\documentclass[acmlarge]{acmart}
\AtBeginDocument{%
  }

\copyrightyear{2026}
\acmYear{2026}
\setcopyright{cc}
\setcctype{by-nc-nd}
\acmConference[FAccT '26]{The 2026 ACM Conference on Fairness, Accountability, and Transparency}{June 25--28, 2026}{Montreal, QC, Canada}
\acmBooktitle{The 2026 ACM Conference on Fairness, Accountability, and Transparency (FAccT '26), June 25--28, 2026, Montreal, QC, Canada}
\acmDOI{10.1145/3805689.3812275}
\acmISBN{979-8-4007-2596-8/2026/06}

\usepackage{graphicx}
\usepackage[dvipsnames]{xcolor}

\newcommand{\TK}[1]{\textcolor{red}{\textbf{TK:} #1}}
\newcommand{\AM}[1]{\textcolor{blue}{\textbf{AM:} #1}}
\newcommand{\JP}[1]{\textcolor{violet}{\textbf{JP:} #1}}

\usepackage[colorinlistoftodos]{todonotes}

\newcommand{\newtext}[1]{#1}

\renewcommand{\TK}[1]{}
\renewcommand{\AM}[1]{}
\renewcommand{\JP}[1]{}

\makeatletter
\newcommand{\myconfshort}{\acmConference@shortname}
\newcommand{\myconffull}{\acmConference@name}
\newcommand{\myconfdate}{\acmConference@date}
\newcommand{\myconfloc}{\acmConference@venue}
\AtBeginDocument{
  \fancypagestyle{firstpagestyle}{
    \fancyhead{}%
    \fancyfoot[C]{}%
  }
  \fancyhf{}
  \fancyhead[LO]{\@headfootfont\shorttitle}%
  \fancyhead[RE]{\@headfootfont\@shortauthors}%
  \fancyhead[LE]{\@headfootfont\footnotesize \myconfshort, \myconfdate, \myconfloc}%
  \fancyhead[RO]{\@headfootfont\footnotesize \myconfshort, \myconfdate, \myconfloc}%
  \fancyfoot[C]{}%
}
\makeatother

\begin{document}

\title{The Racial Character of Computer Graphics Research}

\author{Theodore Kim}
\email{theodore.kim@yale.edu}
\author{Alexa Schor}
\email{alexa.schor@yale.edu}
\author{Julian Posada}
\email{julian.posada@yale.edu}
\author{Alka V. Menon}
\email{alka.menon@yale.edu}
\affiliation{%
  \institution{Yale University}
  \city{New Haven}
  \state{CT}
  \country{USA}
}

\renewcommand{\shortauthors}{Kim et al.}

\begin{abstract}
Computer graphics algorithms for generating photorealistic imagery are widely perceived to be universal, and capable of conjuring anything that a filmmaker or game designer can imagine. However,  recent works have suggested that 3D algorithms for depicting synthetic humans are far from generic, and instead favor historically hegemonic characteristics. We present the first systematic review of human depiction in the top computer graphics conference and the journal of record ({\em SIGGRAPH} and {\em ACM Transactions on Graphics}) that confirms previous hypotheses. Algorithms that claim to be generically rendering ``human skin'' are in fact imagined and  formulated for translucent, ``high albedo'' materials such as white skin. Algorithms claiming to apply generically to ``human hair'' are formulated for ``rods,'' ``wires'' and ``threads'' which are analogous to straight hair. Our analysis reveals conceptual binarization, where algorithms for white skin are treated as computational substrate for ``all'' skin, imposing a hierarchical assumption that all skin descends from the math and physics of white skin. Hair algorithms follow a similar historical pattern, with the first examples of computer-generated Type 4 hair only appearing after the murder of George Floyd in 2020. We offer a new conceptual label, {\em McDaniels Methods}, for characterizing and critiquing computer graphics algorithms that reinforce racial hierarchy under a false cover of diversity. We also offer an inverse label, {\em Durald Methods}, for algorithms that were closely co-designed with the people being depicted. Our analysis points the way towards several neglected avenues for future research.
\end{abstract}

\received{13 January 2026}
\received[revised]{25 March 2026}
\received[accepted]{16 April 2026}

\maketitle

\section{Introduction}
\label{sec:intro}

Modern algorithms for computer generated imagery (CGI) are widely considered to be universal methods that can depict anything in the physical world. Research into CGI has been ongoing for the last 50 years, with a broad constellation of algorithms developed to depict a wide variety of natural phenomena, such as plants \cite{mvech1996visual}, fire \cite{nielsen2022physics}, water \cite{thurey2010multiscale}, muscles \cite{Saito15comp} and cloth \cite{baraff1998large}. Today, the algorithms regularly appear in any form of image-based media, including YouTube videos, film and games. The techniques are sufficiently mature that one VFX practitioner has claimed that ``[t]here’s a general consensus within visual effects that more or less anything can now be simulated realistically'' \cite{warburton2017}. Epic Games has publicly declared \cite{sca2023panel} that the latest version of Unreal Engine overcomes the uncanny valley \cite{mori2012uncanny}. 

However, claims to universalism and neutrality can often mask systematic bias. For example, artificial intelligence systems have been described as ``racialized and gendered models of the self that are falsely presented as universal'' \cite{katz2020artificial} that speak ''to some universal yet never articulated subject when it is obvious that the authors of such systems regard themselves as the gold standard of universal subjects.'' \cite{adam2006artificial} Cryptographic research has seen similar criticisms for ``universalist standard[s] [that] in fact implies a `normal' user who conforms to normative standards of race (White), class (wealthy), gender (cisgendered male), ability (able-bodied).'' The \textit{Gender Shades} study \cite{buolamwini2018gender} punctured such universalist claims by exposing patterns of gender and racial discrimination in supposedly neutral machine learning algorithms. Gender issues have also been identified in the neutrally-coded automatic gender recognition algorithms utilized in HCI \cite{keyes2018misgendering}. Troubling racial patterns have been found in seemingly unbiased restaurant review datasets \cite{luo2024othering}. Rather than being neutral, value systems are implicitly encoded in both datasets and research practices \cite{benjamin2019race,noble2018algorithms,bowker2000sorting,crawford2021excavating,paullada2021data,miceli2022data}. For example, a field-level audit of \cite{birhane2022values} machine learning research found that Performance, Efficiency, and Novelty are vaunted qualities, while harms are rarely discussed. A similar audit \cite{kalluri2025computer} of the computer vision literature found an increasing alignment with surveillance applications, with connections simultaneously being obfuscated.

Similar problems have been found with the claim that CGI algorithms can evenhandedly depict all humans \cite{kim2020racist,kim2022countering}. Key algorithms that are generically billed as capturing the appearance of ``skin'' and ``human skin'' are often only demonstrated on white skin, and the underlying physics being targeted is pale, translucent materials. Algorithms that generically advertise their suitability for depicting the appearance and motion of ``hair'' are often only shown working on straight hair, and investigate the optics and mechanics of infinitely straight rods. These observations are echoed in STS scholarship that observes that, contrary to any claims about universality, CGI algorithms tend to focus on a ``small subset of parts in any given world'' \cite{gaboury2021image}.

Attempts to create CGI humans that deviate from these implicit norms have resulted in Black artists and writers criticizing the poor quality of dark skin and textured hair in video games. \cite{Darke24} \citet{narcisse2017natural} has pointed out that this is partially a problem of representation, i.e.~video game artists with insufficient backgrounds inevitably fail to make authentic-looking hair. However, the biases baked into the underlying CGI techniques also play a significant role. The algorithms were made to depict one kind of human, so creating other types that lie outside the preferred baseline becomes progressively more difficult.

\begin{figure}
\includegraphics[width=0.49 \textwidth]{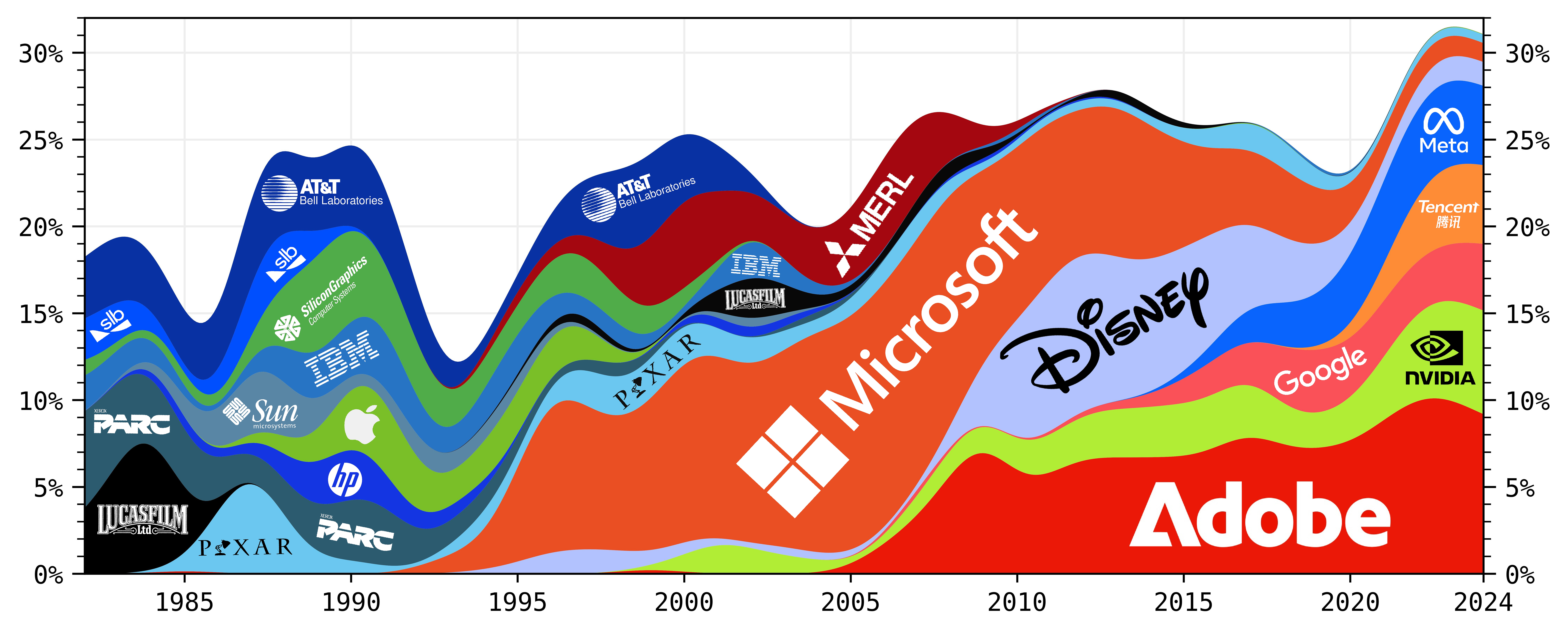} \vrule
\includegraphics[width=0.49 \textwidth]{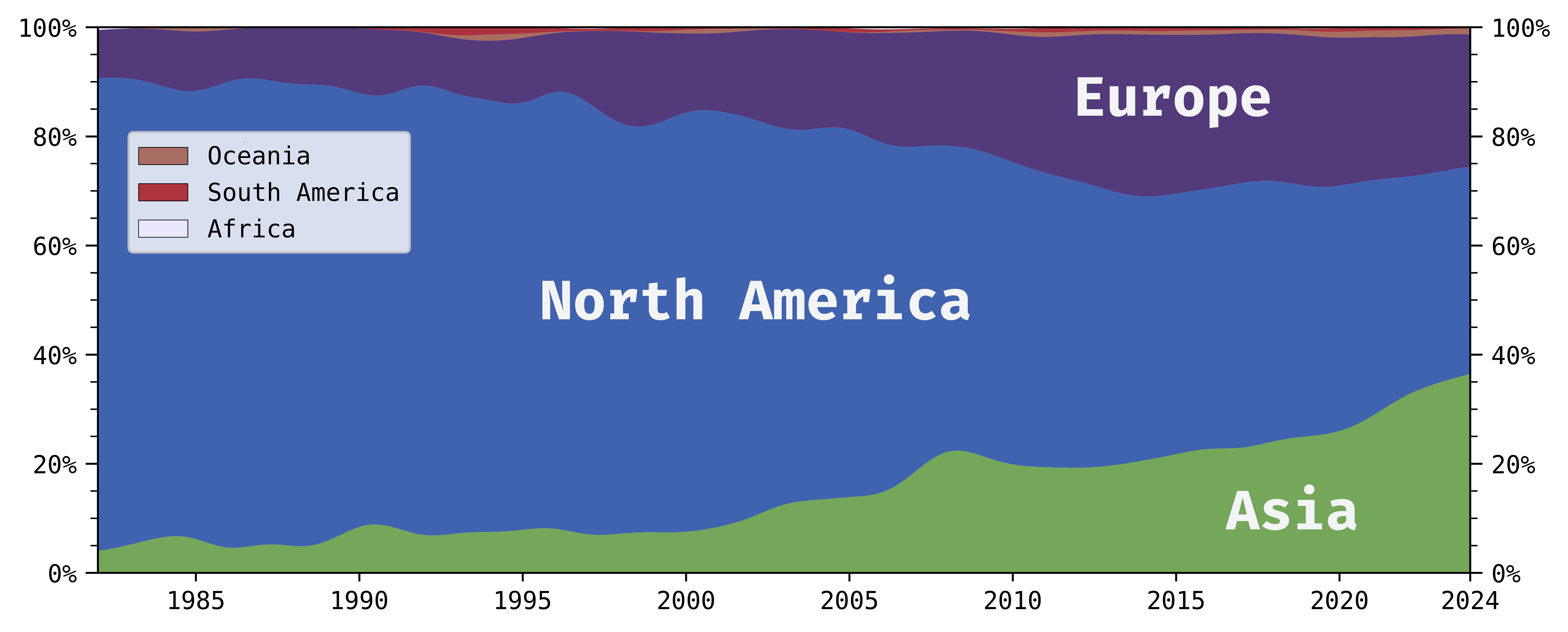}
\caption{\textbf{Left:} Percentage of total papers coming form industrial labs in \textit{SIGGRAPH} and \textit{TOG}. Film studios like Lucasfilm and Pixar had early influence, as well as Disney in the 2000s and 2010s. \textbf{Right:} Geographic location of authors over time. North America dominated early on, and majority currently skews WEIRD (Western, Educated, Industrialized, Rich, and
Democratic). \cite{henrich2010weirdest} The data source for these figures is described in Appendix \ref{appen:data_process}.}
\label{fig:industry}
\end{figure}

We present the first field-level audit of human depiction in computer graphics research papers. We focus on the conferences of record in the field of computer graphics, \textit{SIGGRAPH} (1974-present) and \textit{SIGGRAPH Asia} (2008-present), and the journal of record, \textit{ACM Transactions on Graphics (TOG)} (1982-present). The venues are inextricably linked: for much of their existence, the \textit{SIGGRAPH} proceedings were published as a special issue of \textit{TOG}. Through an agreement with our university library, we have obtained the full text of every article through the end of 2024. The size of this corpus is significant (6864 papers), so we perform a \textit{thematic analysis} of this body of literature in order to build a roadmap for future research. We tag and categorize every instance of the term ``skin'' and ``hair'', identify patterns, and analyze the images and language that are deployed around these terms. A large proportion of the early papers were written by North American and European authors (Figure \ref{fig:industry}, right), so much of what we see is WEIRD (Western, Educated, Industrialized, Rich, and
Democratic) norms \cite{henrich2010weirdest} consistent with the cultural practices of the U.S.~film and video game industries.

Previous works on this topic \newtext{(some written by authors of this paper)} have largely been short form, either speculating \cite{kim2020racist} on the existence of racial patterns in a 1000-word op-ed, or presenting pilot analyses \cite{kim2022countering} in a 2-page extended abstract. We instead present the first detailed, literature-level study that both confirms and extends the hypotheses from these previous works. In addition to white skin being presented as false synecdoche for all skin, we find a community-level binarization where skin is conceptualized as a ``white'' / ``non-white'' dyad. Consistent with similar feminist analyses \cite{klein24data}, the binary encodes a hierarchy that imagines the ``Black body as a derivation from the white body'' \cite{malazita2024enacting}. We find a similar dyad and hierarchy of ``straight'' / ``non-straight'' in the algorithmic conception of hair. We observe an uptick in the instances of Black skin, as well as the \textit{first} instance of Black hair, coincident with the aftermath of the 2020 murder of George Floyd. However, these instances often appear as low-quality addenda to papers that otherwise showcase white skin and straight hair. This practice mirrors early analog cinematography \cite{alton1949painting}, where Black faces were only shown when necessary to suggest that white skin lighting norms were universal.

We propose the term \textit{McDaniels Methods} to identify algorithms that continue this practice of reinforcing racial hierarchy under a false cover of diversity. We define McDaniels Methods as \textit{techniques that are exhaustively validated on white skin and straight hair, and relegate all other forms of skin and hair to a small set of flawed and underwhelming examples.} We choose the name ``McDaniels'' because the one Black face in the influential 1949 cinematography manual {\em Painting With Light} \cite{alton1949painting} is that of Hattie McDaniel, the first Black actor to ever win an Academy Award in 1939. Reflecting the carelessness her depiction was given at the time, the book misspells her name. The caption lists her as \textit{Hattie McDaniel\textbf{s}}, but her last name has no trailing ``s''. We call such techniques \textit{McDaniels Methods} to underscore how the racial practices of the analog era carried forward into the digital.

\newtext{We propose the contrasting term \textit{Durald Methods}, which we define as \textit{techniques that were co-created in close collaboration with the people being depicted.} We name the term after Autumn Durald Arkapaw, the first woman of color to ever win the Academy Award in Best Cinematography for the 2025 film \textit{Sinners}.}

\section{Background} 
\label{sec:background}

Race is a sociopolitical construct with many dimensions. From self-identification, to phenotype, to evaluations of racial group membership by others on the basis of appearance to ancestry or genetics, the word ``race'' is colloquially used to cover all these different dimensions \cite{roth2016multiple}. In this paper, when we discuss the ``racial character'' of computer graphics, we draw on the dimension of race as represented through phenotype, including color and physical features like hair texture. Phenotype encompasses the visual dimension of race, which is key for representational contexts. Its mapping onto racial categories is historically and culturally contingent, and central to the construction and maintenance of racial hierarchies in the U.S. \cite{hammonds2013new,omi2014racial}. This then echoes in international contexts and online \cite{nakamura2007digitizing} because of the U.S.’s economic and cultural power \cite{menon2023refashioning}. Attempts to represent race reveal the assumptions a society makes about bodies, values, and universality. Fields from cosmetics to medicine have sought to render the social phenomenon of race exclusively in technical and physical terms, under the guise of colorblindness \cite{bonilla_silva_2021}. The principle of colorblindness ``pretends that racial recognition rather than racist rule is the problem to be solved'' \cite{crenshaw2019seeing} and suppresses inquiry by ``equat[ing] \textit{noticing} racial inequality with \textit{perpetuating} racial inequality'' \cite{ray2023critical}.

In the absence of such inquiries, a norm of whiteness becomes the default, the unmarked category \cite{menon2017reconstructing,sexton2008amalgamation}. This stems from the enduring legacy U.S. and European domination, where white people topped the racial hierarchy \cite{morning2011nature} and came to stand in for the universal human \cite{epstein2024inclusion}. At the same time, scholars have pointed to a preference for lightness in Africa and Asia that can overlap with and reinforce norms of whiteness, even as they arise from locally specific social hierarchies \cite{glenn2008yearning,thomas2020beneath}. Thus, representation of race, even without any explicit references, remakes and produces racial difference even as it purports to simply describe it.

This is not a new phenomenon, as whiteness as the default in imaging technologies goes back to the analog era. \citet{roth2009looking} investigated its role as a baseline in analog color photography, and \citet{yue2020girl} examined how it intersected with gender when looking at ``Leader Ladies'' or ``China Girls'' as development targets in analog color motion pictures. \citet{dyer1997white} traces the provenance of these norms back through silent film era to the Dutch Old Masters. Extending these analyses, our work shows how the same racial standards and practices  carried over from the analog into the digital era.

This research advances the wider critique of colorblind racism and the universalism applied to computing technology. Our analysis is notably informed by the ``New Jim Code'' of \citet{benjamin2019race}, which critiques technological deployments portrayed as objective or progressive that actually reproduce existing racial inequalities. This resonates with the work of \citet{noble2018algorithms} on algorithmic oppression, which challenges the perceived neutrality of computer-mediated platforms that allow structural inequality to propagate under a colorblind veneer. These authors extend the critique of colorblind racism leveled by \citet{bonilla_silva_2021}, specifically the naturalization of racial outcomes and the abstract liberalism used to justify market dynamics, which ultimately renders structural inequalities invisible. Such masking of social differentiation has been a persistent feature of computing technology for decades, as scholars continue to document how the “myth of neutrality” sustains systemic bias.

In this context, many have examined the values encoded in the research practices of different sub-disciplines in computer science. Cryptography \cite{rogaway2015moral}, machine learning \cite{birhane2022values}, computer vision \cite{kalluri2025computer}, and human-computer interaction \cite{dourish2006implications} have all seen thematic analyses of what is implicitly valued and devalued in their respective literatures. Rather than reflecting neutral, disinterested research that is taking place from ``a view from nowhere'' \cite{adam2006artificial}, many of these sub-disciplines are repeating and re-entrenching the biases and power structures of the past. For example, some machine learning datasets have been found to be overwhelmingly light-skinned \cite{buolamwini2018gender}, which produces facial analysis algorithms that perform poorly on darker-skinned females, and has led to a variety of works on bias in visual datasets \cite{fabbrizzi2022survey}.

While thematic analyses have been carried out most prominently on machine‑learning datasets and their outputs, far less attention has being paid to computer graphics algorithms. Such analyses are urgent because CGI algorithms are increasingly being used to generate ``synthetic'' datasets, for example, for machine learning. Notable examples include driving simulation \cite{dosovitskiy2017carla}, human faces \cite{wood2021fake}, and full human bodies \cite{varol2017learning}. In this sense, our analysis lies one step upstream: any biases present in the underlying algorithmic substrate will be passed on through the data. 

Science and Technology Studies (STS) scholars have also begun examining the bias in CGI algorithms. Building on earlier analyses \cite{kim2020racist}, \citet{malazita2024enacting}  examines ``white photorealism'' in the Unreal Engine, as \citet{gass2022body} has similarly shown in the 2018 game \textit{Detroit: Becoming Human}. Other work has shown how a normative white face is the baseline for \textit{FaceGen}, the character creation tool used in \textit{Fallout 3}. \citet{phillips2020gamer} and \citet{ketchum2009facegen}  have shown how CGI algorithms are non-universal, and how reality is "unevenly distributed" in synthetic images \cite{gaboury2021image}. STS scholars \cite{dodik2022sex,malazita2022using} have collaborated with computer graphics researchers on extended abstracts at \textit{SIGGRAPH} that bring more visibility to these issues. None of this previous work has attempted a field-level audit of representation. Thus, our work both provides empirical validation and extension to this prior scholarship.

\paragraph {Conference and Journal Choice}
We selected \textit{SIGGRAPH} and \textit{TOG} for analysis because they are widely acknowledged as the leading venues in computer graphics. The annual \textit{SIGGRAPH} conference routinely hosts over 10,000 attendees \cite{CGW2015}, including leading researchers from both industry and academia, and has historically been where major breakthroughs appear. The next largest \textit{Eurographics} conference is over an order of magnitude smaller at roughly 500 annual attendees \cite{Euro14}. \textit{TOG} is consistently the top-rated computer graphics journal, as evidenced by 2024 Google Scholar Metrics (h-index 114), Journal Citation Reports (Impact Factor 9.5), and Eigenfactor (0.031). The next highest journals trail significantly: \textit{Computer Graphics Forum} (h-index 64, IF 2.9, EF 0.023) and \textit{IEEE Transactions on Visualization and Computer Graphics} (h-index 94, IF 6.5, EF 0.017). The latter's citations mostly come from data visualization, not computer graphics. 
While a complete analysis of all computer graphics papers would undoubtedly yield valuable insights, any attempt to understand what is considered \textit{top-quality, high-prestige} research in computer graphics must start by examining the papers in \textit{SIGGRAPH} and \textit{TOG}.

\paragraph{Skin Classification Systems}
Numerous schemes have been developed, the most prevalent being the Fitzpatrick scale \cite{fitzpatrick1975soleil} that classifies skin based on how quickly in burns in the sun. Fitzpatrick \cite{fitzpatrick1988validity} disputed this application of his system, because it had been developed ``to classify persons \textit{with white skin}'' (emphasis his). The von Luschan \cite{von1897beitrage} scale was developed for racial classification, but was ``swiftly abandoned when reflectance spectrophotometry was introduced in the early 1950s'' \cite{jablonski2004evolution}. Its role in Nazi eugenics \cite{vonLuschan2017} probably contributed to this abandonment. The recent Monk \cite{monk2023monk} scale adopted by Google and Meta is an effort to decouple ``tone'' from ``race.''

Classifying skin is inherently fraught. The concept of ``white'' varies widely by geographic region and political context because ``whiteness can only be defined by state power'' \cite{cottom2019thick}. 
That is, country-specific laws, like the U.S. justice system, ``adjudicate the arbitrary inclusion and exclusion of people across time,'' \cite{cottom2019thick} such as with Irish American immigrants in the 19th century \cite{ignatiev2012irish}. After extensive discussion of the constraints of existing systems, the authors decided on the broad classifications of ``white'' and ``non-white.'' These labels are designed to facilitate the overarching purpose of our study: a conversation about what has remained implicit in the field.

\paragraph{Hair Classification} Many hair typing systems also exist with various drawbacks. For example, the popular Walker \cite{Walker1997} system (Types 1-4) is a qualitative measure developed for haircare products, and does not correspond to physical measurements. The system has been criticized for excluding curlier hair types \cite{CurlCentric}, leading to the informal addition of Types 3c and 4c. L'oreal developed a quantitative system \cite{de2007shape} that measures the ``number of waves'' when a hair is pressed between two plates of glass, a process that breaks down with helical hairs.

We adopted a qualitative approach to hair types that is a reduced version of the Walker \cite{Walker1997} system:
\begin{itemize}
    \item \textit{Straight and wavy hair}, containing by virtually no waves, or roughly sinusoidal waves along one dimension. This hair does \textit{not} contain full helical twists.
    \item \textit{Curly hair}, containing full helical twists, but with a sufficiently large radii that hair is straight near the scalp.
    \item \textit{Textured hair}, which contains twists with millimeter-scale radii, and forms a spongy layer near the scalp.
\end{itemize}
In the Walker system, straight (Type 1) and wavy (Type 2) hair are separate categories. We combined them because it was often unclear if an undulation was inherent to the hair or originated from styling, e.g.~because the hair had been lightly mussed.
The co-authors with the computer graphics background performed the initial classifications of skin and hair in the articles, which were spot checked by the sociology authors.

\section{Methods}

\paragraph{Skin Papers}
\label{sec:skin_methods}

We initially searched for all \textit{SIGGRAPH} and \textit{TOG} papers containing the word ``skin'' that were also tagged under the ACM Computing Classification System (CCS) as \textbf{Computing Methodologies $\rightarrow$ Computer Graphics $\rightarrow$ Rendering}. Since \textit{SIGGRAPH} began in 1974 but the current CCS system was instituted in 2012, the tags were unreliable. For example, one influential paper \cite{jensen2001practical} was incorrectly tagged as  ``shape modeling'' and ``computational geometry''. 

Instead, we searched the corpus with the query \texttt{skin AND (render OR rendering OR shading)} to obtain a superset of the relevant papers. The query yielded 466 papers. We manually inspected the results, and if the term ``skin'' referenced a photorealistic rendering in the paper, it was included. Second, if the term ``skin'' was used in the text to refer to the physics of skin interacting with light, the paper was included. 120 papers remained after this culling process. Table \ref{tbl_skin_exclude} gives a breakdown of the excluded papers.

The authors coded skin color as white/non-white in the images appearing in the selected papers. We chose not to outsource the coding process to crowd-based ``data workers'' because prior research has documented concerns with such practices, including the risk of client-induced bias in annotated outputs \cite{miceli2022data} and limited transparency surrounding the labor conditions and demographic composition of workers on major platforms \cite{muldoon2025poverty}. Instead, two authors who are North American computer graphics researchers performed initial classifications, which were then spot checked by the sociology authors. The labels from the two researchers only disagreed on 3 papers \cite{yoshihiro2018relighting,bickel2007multi,nishino2004eyes} out of 120. The disagreements were resolved through discussion, detailed in Appendix \ref{append:skin}.
Since the field has been long-dominated by North American scientists (Figure \ref{fig:industry}), collecting labels from researchers embedded in that population provides the most direct window into the field's perceptions. The labels are not authoritative but subjective, and rooted in the researchers' commonsense understanding. 
\begin{table}
\begin{tabular}{ |p{1cm}||p{13cm}|  }
 \hline
Number & Category \\
 \hline
146   & No image of a human or substantive reference to skin, e.g.~``skin'' only appeared in a citation.   \\
39 & Image and video editing papers. These 2D algorithms are distinct from 3D rendering methods. Papers spanning 2D and 3D, e.g.~using 2D images to construct a 3D avatar, remained included.\\
37 & Incidental image of humans, e.g.~a geometry processing paper containing a human head model \cite{tasdizen2003geometric}\\
16 & Hair papers, moved to our hair analysis.\\
21 & Geometric skinning, elasticity, and motion capture papers on skin \textit{animation}, not rendering.\\
17 & Eyelid \cite{bermano2015detailed}, and eyebrows \cite{li2023ems}, teeth, tongue, and lips papers.\\
14 & Full-body capture papers, as issues of body normativity are out-of-scope.\\
13 & Clothing capture and simulation papers\\
20 & Anatomy modeling, e.g.~muscles \cite{saito2015computational}, diaphragm \cite{dilorenzo2008laughing}, and hands.\\
13 & Illustration papers, e.g.~portrait painting tools \cite{haeberli1990paint} or anime portrait generators \cite{li2023parsing}.\\
5 & Medical imaging papers, e.g.~viewing organ shapes in MRI data.\\
5 & Haptic rendering \cite{long2014rendering} and crowd simulation papers\\
\hline
\end{tabular}
\caption{Papers excluded from our analysis from the query \texttt{skin AND (render OR rendering OR shading)}.}
\label{tbl_skin_exclude}
\end{table}

\paragraph{Hair Papers}
\label{sec:hair_methods}

We performed the query \texttt{hair AND (render OR rendering OR shading OR modeling OR modelling OR simulation OR animation)}, yielding 570 papers. We manually inspected the matches, and if the term ``hair'' appeared in the context of a photorealistic image, such as an attempt to achieve realistic appearance, shape, or motion, the paper was included. If no image appeared, but `hair' was in reference to the optics or mechanics of hair, it was included. 160 papers  remained after this culling process. Table \ref{tbl_hair_exclude} gives a breakdown of the excluded papers. Two graphics researchers coded hair as straight/wavy, curly, or textured, which were then checked by the sociology authors. Only 3 papers of 160 \cite{heitz2015sggx,jakob2009capturing,wang2020single} saw disagreement, and were resolved through discussions described in Appendix \ref{append:hair}.

\begin{table}
\begin{tabular}{ |p{1cm}||p{13cm}|  }
 \hline
Number & Category \\
 \hline
130   & No image of hair or substantive reference to hair, e.g.~``hair'' only appeared in a citation.   \\
67 & Image and video editing papers. These 2D algorithms are distinct from 3D rendering methods. Papers spanning 2D and 3D remained included.\\
51 & Illustration and stylization papers, e.g.~watercolors \cite{chu2005moxi}, pen and ink \cite{salisbury1997orientable}, and manga \cite{xie2020manga}. \\
63 & Face-only papers, that capture and render faces, with hair excluded \cite{guenter1998making,bradley2010high,wu2018deep}. \\
42 & Full-body and clothing capture papers where hair detail is coarse \cite{de2008performance} or removed \cite{liu2023towards}. \\
22 & Geometry papers, referencing ``hair'' region on geometric model, e.g.~Michaelangelo's \textit{David}.\\
24 & Incidental image of humans, e.g.~differentiable rendering where one of the examples is a head. \cite{laine2020modular}\\
7 & Eyelash, eyebrow, and feather papers.\\
4 & Crowd simulation and perception papers.\\
\hline
\end{tabular}
\caption{Papers excluded from our analysis from the query \texttt{hair AND (render OR rendering OR shading OR modeling OR modelling OR simulation OR animation).}}
\label{tbl_hair_exclude}
\end{table}

\paragraph{Limitations}

Other computer graphics conferences include \textit{Eurographics}, \textit{Pacific Graphics}, and \textit{Afrigraph} respectively based in Europe, Asia, and Africa. Preliminary examinations of these corpora did not significantly shift our findings. Strikingly, we did {\em not} find a sudden wealth of Black skin and textured hair papers within \textit{Afrigraph}. Instead, both the African and European conferences contained the same patterns of showcasing white skin \cite{Struck04real,d2007efficient} and straight hair \cite{patrick03lightwave,huang22micro}, and we were able to locate only two instances of textured hair \cite{patrick2004,bertails2005}. While the European (1980 - present) and Asian (1993 - present) conferences are long-running, the African conference was short-lived (2001 - 2010).

Our choice of classifications limits the patterns we can identify. Hair color was not categorized, and skin tone was not evaluated on a continuum. Elective modifications such as the straightened textured hair \cite{hooks1989talking}, lightened skin \cite{thompson2015shine}, dyed hair, or tanned skin were not tagged. Potential trends such as East Asian skin representations becoming lighter (or darker) over time, Black subjects being shown predominantly with straightened hair, or blonde hair being preferred in general, cannot be detected using our labels, and are interesting avenues for future inquiry. Sub-categories of ``non-white'' like East Asian and South Asian were coded, but did not alter the overall findings. Additional observations from these labels are in Appendix \ref{appen:results}. 

This reflexive thematic analysis \cite{braun2006using} is necessarily a subjective and messy exercise. Outcomes may vary depending on the person performing it and the assumptions and racial commonsense they bring. Again, our point is not to establish an authoritative, replicable classification system, nor to use an existing system given the significant problems each has. Rather we seek to characterize the implicit assumptions in the computer graphics field in order to lay the groundwork for a more explicit conversation about racial representation going forward.

\section{Results}

\subsection{Skin Papers}

Our survey of the literature suggests that the term ``skin'' takes on at least three different meanings within the context of rendering. In order to understand their relationship, we will examine them in order of smallest to largest spatial scale:
\begin{itemize}
    \item Skin as an \textit{optical substrate}: the physics of light propagation through an infinitesimal physical medium. E.g.~how a laser beam spreads as it impinges on a piece of marble \cite{goesele2004disco}.
    \item Skin as a \textit{layered material}: light propagation through a mesoscale, heterogeneous medium. E.g.~how light interacts with the oils and melanin on a swatch of skin, as well as the hemoglobin (blood) in deeper layers.
    \item Skin as a \textit{human face}: characterizing light transport on a macro object like a head or body. This broadest category contains {\em facial capture}, which tries to acquire the appearance of a human subject's face, and {\em reflectance field capture}, which tries to acquire the appearance of generic objects, including humans.
\end{itemize}
For all the following findings, more detailed breakdowns are available in Appendix \ref{appen:results}.

\begin{figure}
\includegraphics[width=\textwidth]{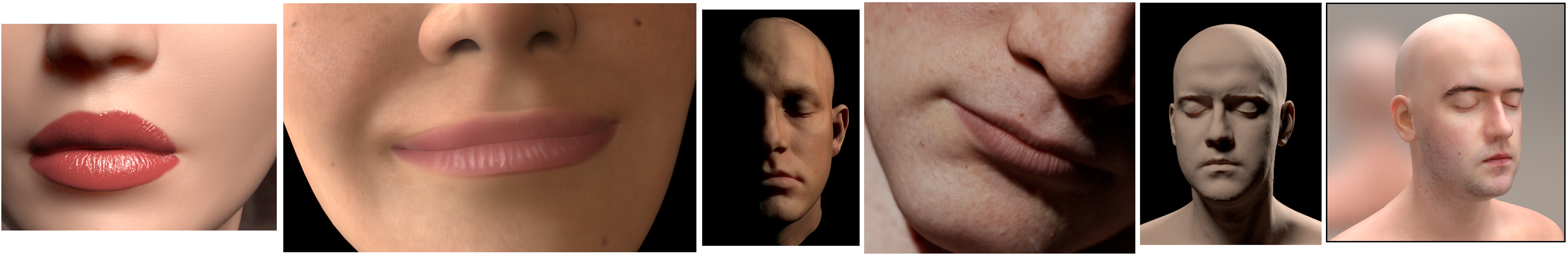}
\caption{3D renderings from papers that investigate skin as an optical medium, respectively \cite{jensen2001practical,jensen2002rapid,donner2005light,d2011quantized,frederickx2017forward,vicini2019learned}, from {\em SIGGRAPH}, {\em SIGGRAPH Asia} and {\em ACM TOG}. There are no other renderings of humans in these papers.}
\label{fig:white_skin}
\end{figure}

\subsubsection{Skin as an optical medium}
\label{sec:skin_optical}

The 19 papers in this category suggest that at the basic optical level, \textit{SIGGRAPH} researchers imagine skin as a pale, translucent material. Papers at this level repeatedly state equivalences to milk, marble, and snow. The early 1993 work of \citet{hanrahan1993reflection} established this equivalence by grouping together ``natural materials such as \textbf{human skin}, plant leaves, \textbf{snow}, sand, paint''. As a result, \textit{subsurface scattering}, the physics of how light spreads diffusely through space as it propagates through a translucent material, dominates the research at this spatial scale. Subsequent to the 1993 paper \cite{hanrahan1993reflection}, follow-on work goes further and draws equivalences to other extremely pale materials:
\begin{itemize}
\item ``translucent materials, such as \textbf{marble}, cloth, paper, \textbf{skin, milk, cheese}, bread, meat, fruits, plants, fish, ocean water, \textbf{snow}, etc.'' \cite{jensen2001practical}
\item ``Translucent materials are frequently encountered in the natural
world. Examples include \textbf{snow}, plants, \textbf{milk}, \textbf{cheese}, meat, \textbf{human skin}, cloth, \textbf{marble}, and jade'' \cite{jensen2002rapid}
\item ``The accurate and efficient simulation of these effects is often
required to achieve the color and soft appearance of media such as
\textbf{skin}, hair, ocean water, \textbf{wax} and \textbf{marble}.'' \cite{d2011quantized}
\item ``common materials are partially permeable to light such as \textbf{wax}, \textbf{skin}, teeth, \textbf{cheese}, fish, or stone'' \cite{brunton20183d}
\end{itemize}
With the exception of the 1993 paper, \cite{hanrahan1993reflection} all of the subsurface scattering papers  \cite{jensen2001practical,jensen2002rapid,donner2005light,d2011quantized,frederickx2017forward,vicini2019learned} only show renderings of ``skin'' that feature pale, white skin (Figure \ref{fig:white_skin}). There are no images of Black (high melanin) skin, and the aesthetics and optical phenomena associated with high-melanin skin are not considered. 

\begin{figure}
\includegraphics[width=\textwidth]{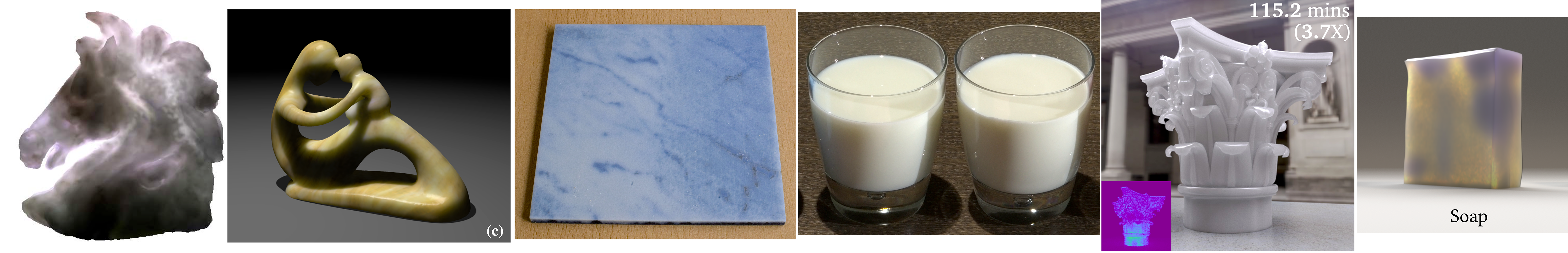}
\caption{Examples from other papers presenting optical materials that are deemed analogous to ``skin''. Alabaster \cite{goesele2004disco}, jade and wax \cite{song2009subedit}, translucent plastics \cite{havsan2010physical}, milk \cite{papas2013fabricating}, marble \cite{zhao2014high}, and soap \cite{deng2022reconstructing}. No humans appear in these publications.
}
\label{fig:translucency}
\end{figure}

Many subsurface scattering papers do not contain any renderings of humans and focus on e.g.~jade figurines or slabs of marble. They still consistently draw equivalences between ``skin'', subsurface scattering, and pale materials. The language is consistent across papers, decades, and research groups, suggesting community-wide consensus. (Figure \ref{fig:translucency})
\begin{itemize}
\item ``Many daily life objects (e.g., \textbf{milk, skin or marble}) are translucent'' \cite{goesele2004disco}
\item ``important for simulating the appearance
of translucent materials, such as \textbf{marble, skin, and milk}.'' \cite{hao2004real}
\item ``materials such as \textbf{wax, marble, and skin} exhibit
light scattering within the object volume.'' \cite{song2009subedit}
\item ``subsurface scattering, which is important for translucent materials such as \textbf{wax, marble,} food, and \textbf{skin}.'' \cite{dong2010fabricating}
\item ``this scattering is crucial to believable appearance of materials
such as human \textbf{skin, wax, and marble}.'' \cite{havsan2010physical}
\item ``thick media (such as \textbf{soap, skin} and jade)'' \cite{huo2016adaptive}
\end{itemize}
The above papers were all written after 2001, after the publication of a highly influential 2001 ``dipole approximation'' paper \cite{jensen2001practical} that saw widespread commercial adoption. The algorithm was used to render the translucent characters of Gollum in {\em The Two Towers} (2002) \cite{aitken2004lord}, Dobby in {\em Harry Potter and the Chamber of Secrets} (2001) \cite{hery2004rendering}, and won a 2004 Scientific and Technical Academy Award. It currently has 1384 citations on Google Scholar, exceeding the citation count of its predecessors (803 and 494 citations for \cite{hanrahan1993reflection} and \cite{marschner1999image}).

\AM{Not sure this is directly speaking to your point, but this adoption is against a longstanding, global preference for “lightness” and “brightness,” if not outright whiteness (Nakano Glenn 2008; Hunter). An implicit racism means that Blackness is not even considered as something to be achieved for people (as subjects) in the aesthetic realm. There is a default to whiteness as representing the “standard” human (Sexton? Lipsitz?).}

The equivalence of skin and snow is both implied and directly stated in commercial deployments. A 2018 special issue of {\em TOG} published full-length invited papers from all the major commercial and film studio rendering teams. Of these, the papers for Pixar's \textit{RenderMan}, Walt Disney Animation's \textit{Hyperion}, and Weta Digital's \textit{Manuka} rendering engines appeared in our keyword search. All three only label renderings of white, pale faces as instances of ``skin,'' and the \textit{Hyperion} paper \cite{burley2018design} even describes how their approach to skin evolved from their code for rendering snow.

\begin{figure}
\includegraphics[width=\textwidth]{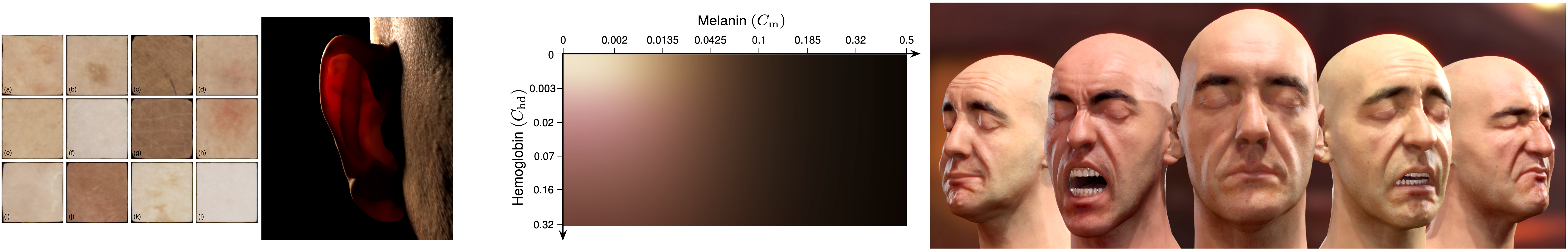}
\caption{Examples from papers presenting layered models of skin. In the left pair, swatches of different skin color are presented, but the final renders only show white skin. \cite{donner2008layered} In the right pair, a skin color lookup texture is mentioned, but the final results are only shown on white skin. \cite{jimenez2010practical}.}
\label{fig:layers}
\end{figure}
\subsubsection{Skin as a layered material}

Skin is composed of multiple layers (stratum corneum, epidermis, and dermis \cite{chen2015hyperspectral}) and with the optical media abstraction of skin established in the 1990s and 2000s, subsequent papers attempted to model and render these heterogeneous layers (see Figure \ref{fig:layers}).

Throughout the 2000s and 2010s, layered material papers routinely showed only renderings of white skin as evidence of algorithmic effectiveness \cite{donner2005light,chen2015hyperspectral,nagano2015skin}. In one instance, the algorithm was based on data collected from ``one Caucasian female, 33 years old; three Caucasian males, 26, 33, and 35 years old'' \cite{jimenez2010practical}. In multiple instances, the existence of darker skin is acknowledged in the form of a ``African, skin type V'' swatch \cite{donner2008layered}, a ``skin color lookup texture'' \cite{jimenez2010practical} and a model for melanin variation \cite{chen2015hyperspectral}. \textit{None} of these papers showed final renders of anything except white hands and faces.

\citet{weyrich2006analysis} from 2006 is worth closer examination, as one of the few papers that renders a Black face. The paper collects data from 149 subjects across a wide range of race, gender, and age. One of the main contributions is a handheld instrument for measuring translucency values, but the measurements show that the effect is \textit{significantly} less pronounced in darker skin (Figures 11 and 12 in \cite{weyrich2006analysis}). Black skin is not a block of wax, but as subsurface scattering becomes less pronounced, the possibility that a different visual feature might come to the fore is not explored.

However, it is well-known that the {\em gloss} and {\em shine} on darker skin, i.e.~sharp and bright glints, are the more important visual cues. Many Black cinematographers have emphasized \cite{lynton1988school,harding2017keeping,latif2017s} that these features better convey shape and performance \cite{wiggum2021pixar} with dark skin. Nevertheless, \citet{weyrich2006analysis} does not posit the need for a better models for shine on skin, and instead re-uses established models for metals \cite{torrance1967theory} and plastics \cite{blinn1982light}. One explanation for this is that existing models were sufficiently mature that new works were not necessary, i.e.~the problem of glossiness was already solved. Later papers contradict this hypothesis. Starting in 2014, several papers performed detailed investigations into the optics of layered, glossy materials, and consistently acknowledged its importance in skin appearance:
\begin{itemize}
    \item ``layered biological structures like leaves, flower petals, or \textbf{skin}'' \cite{jakob2014comprehensive}
    \item ``Previous works on layered materials ... such as \textbf{human skin}'' \cite{zeltner2018layer}
    \item ``Hanrahan and Krueger [1993] proposed a general model for layered materials ... Stam [2001] generalized Hanrahan and Krueger’s work taking into account rough boundaries, in the context of \textbf{skin rendering}.'' \cite{guillen2020general}
    \item ``many approximate models targeting specific layered materials
such as ... \textbf{human skin}'' \cite{guo2023metalayer}
\end{itemize}
None of these papers render a glossy-skinned human, Black or otherwise. The rendered objects are instead consistently household items like vases \cite{guo2023metalayer,jakob2014comprehensive}, espresso machines \cite{zeltner2018layer}, and pearlescent soap dispensers \cite{guillen2020general}. We discuss this reluctance in \S\ref{sec:discussion}, relative to Thompson's notion of the ``slave sublime'' \cite{thompson2015shine}. Finally, we found one paper \cite{vilesov2022blending} on the effects of ``darker skin tones'' on blood flow monitoring and one \cite{piovarci2023skin} on tattooing across multiple skin tones. Both appeared post-George Floyd, which we will also discuss in \S\ref{sec:post_floyd}.

\subsubsection{Skin as a Human Face}
\label{sec:skin_face}

As the largest category of ``skin,'' we sorted these papers into two broad sub-categories:  {\em facial capture} and {\em reflectance field capture}. Since both dealt with the rendering humans on the scale of a face or body, the categories contained overlaps. All these papers used the \textit{optical substrate} and \textit{layered material} definitions of ``skin'' as both computational and epistemic baselines. We examined their racial distribution of examples to understand what was deemed sufficient evidence that an algorithm worked on ``humans.''

\paragraph{Facial Capture}

We examined 67 papers on facial capture, animation and rendering. For inclusion in this subcategory, a paper's goal must be the generation of a photorealistic human face from 3D data. This included taking 2D photos, generating a textured 3D model, and then rendering with novel viewpoints, lighting, and/or facial expressions. 

Within this set, 27\% of the papers (18 of 67) contained only white examples. The papers are not localized in time, and span the 1980s \cite{waters1987muscle}, 1990s \cite{guenter1998making}, and 2010s \cite{thies2015real}, with the most recent appearing in 2023 \cite{aneja2023clipface}. The remaining papers all contained non-white examples, but their distribution is highly uneven. Overall, 28\% (10 of 67) of the papers contained Asian, but no Black, examples, while 4\% (3 of 67) contained a only single Black example: Barack Obama.

Among the 39\% of papers (26 of 61) that did contain other Black examples, almost all (21 of the 26) were published after 2020, again the post-George Floyd pattern we will discuss further in \S\ref{sec:post_floyd}. Finally, 4\% of the papers (3 of 67) feature \textit{only} light-skinned East Asians as examples, and all appeared after 2020. We discuss this instance of racial elasticity further in \S\ref{sec:elasticity}.
Under the current \textit{SIGGRAPH} norms, it is still acceptable to publish a paper containing solely pale-skinned (white \cite{aneja2023clipface} or East Asian \cite{chen2024monogaussianavatar}) examples as evidence that an algorithm can render ``humans.''

\paragraph{Facial Reflectance and Relighting}

We examined 19 papers on the topic of facial reflectance field capture, whose initial application was the problem of ``relighting'' in a Hollywood films, i.e.~changing the lighting on an actor's face in post-production. Early versions of this technology as used in, e.g. \textit{Spiderman 2} (2004) \cite{Restuccio2004}.

Within this set, we found 3 papers (16\%) contained white-only final renders, with the most recent published in 2024 \cite{li2024uravatar}. The rest contained non-white examples, but their distribution was again highly uneven. Asians were the only non-white examples in 3 papers (16\%), one contained a single Latina example \cite{guo2019relightables}, and of the remaining 12 papers (63\%), the vast majority (9 of the 12) were published after 2020.

\begin{figure}
\includegraphics[width=\textwidth]{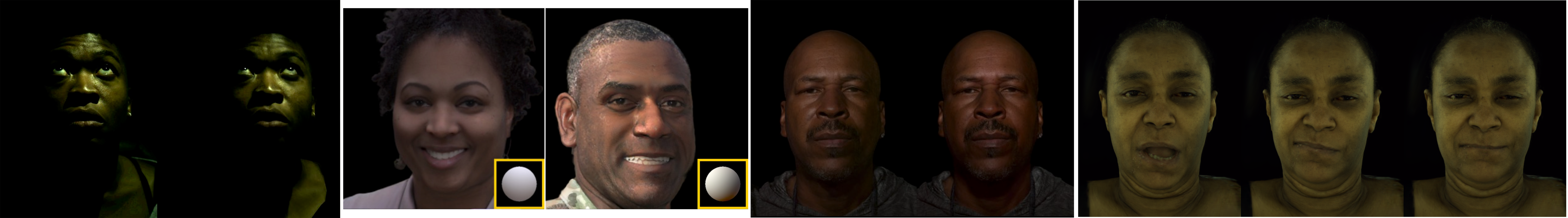}
\caption{Examples of Black skin in relighting algorithms, respectively \cite{bi2021deep,yeh2022learning,sarkar2023litnerf,yang2024vrmm}. The skin is consistently underlit (too dark to see details), and sharp, well-defined highlights (``gloss'') characteristic of dark skin are often muted or missing.}
\label{fig:black_relighting}
\end{figure}
In the papers showing a range of tones, the Black examples usually appear at the end, after the algorithm has been extensively presented and verified on white examples. They are also consistently flat and underlit (Figure \ref{fig:black_relighting}), suggesting that the input photographs were taken under the same lighting setups as the white examples, and the outputs validated under the same conditions. The low contrast and flat colors recall the 2020 controversy \cite{Thompson2020,Schild2020} surrounding a \textit{Vogue} \cite{Aguirre2020} article where white photographer Annie Leibovitz shot  underlit and unflattering images of Black gymnast Simone Biles. 
The relighting papers similarly display Black skin under white lighting.

\subsection{Hair Results}

Whereas the computer graphics conception of ``skin'' is most commonly associated with its appearance, research into ``hair'' can reference its appearance, shape or motion. We decomposed hair papers into three categories:
\begin{itemize}
    \item \textit{Geometry} methods for constructing, editing, and capturing the \textit{shape} of hair.
    \item \textit{Simulation} methods for computing the \textit{motion} of hair.
    \item \textit{Rendering} methods for computing the \textit{appearance} of hair, particularly how light interacting with hair.
\end{itemize}
These broad topic categories already exist in the \textit{SIGGRAPH} literature, so we are largely following the categorizations that the papers self-report.

\begin{figure}
\includegraphics[width=\textwidth]{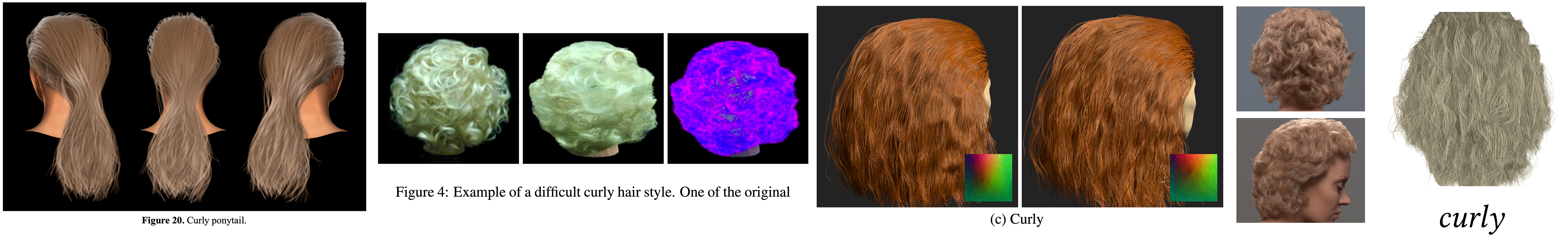}
\caption{Inconsistent labels for ``curly'' hair. From left to right, ``curly ponytail'' \cite{kim02interactive}, ``curly hair'' \cite{wei2005modeling}, ''curly'' \cite{wang2009example}, ``curls'' \cite{herrera2012lighting} and ``curly'' \cite{Reshetov24modeling}. Many only contain light waves, and do not contain a full helical revolution (i.e.~curl).}
\label{fig:curly_hair}
\end{figure}

\paragraph{Hair Geometry}

We identified 61 hair geometry papers. Of these, 27 papers (44\%) showed results exclusively on straight or wavy hair, the most recent from 2024 \cite{buehler2024cafca}. Two additional papers (3\%) showed only straight and wavy examples, plus Barack Obama. Another two (3\%) dealt with ``braids,'' which meant straight hair braided into ponytails or pigtails. Braided textured hair styles such as box braids or cornrows were not considered.

We found 26 papers (43\%) that showed ``curly'' examples in addition to straight and wavy examples. However, the definition of ``curly'' was \textit{highly inconsistent} across the literature, and sometimes corresponded to hair that our computer graphics author labeled as ``wavy.'' They did not contain anything approaching a full helical revolution (Figure \ref{fig:curly_hair}). Textured hair appeared in 4 papers (7\%), two of which were as failure cases, where the paper's algorithm was shown to fail on box braids  \cite{saito20183d} and ``frizzy'' hair \cite{chai2015high}. Only one 2024 paper \cite{wu24curly} explicitly attempted to model textured hair.

\paragraph{Hair Simulation}

We identified 28 simulation papers that contained images of hair. Of these, 19 papers (68\%) exclusively showed straight or wavy hair, while 5 (18\%) papers also included curly hair. Two papers \cite{hsu2023,crespel2024} included textured hair, and were respectively from 2023 and 2024.

Another 7 simulation papers did not show images of hair, but made reference to ``hair'' that illustrated how the concept is situated in the scientific imagination:
\begin{itemize}
    \item ``physical objects such as \textbf{pieces of rope and strands of hair}'' \cite{miller1988motion}
    \item ``\textbf{hair, chains,} and other articulated models'' \cite{redon2005adaptive}
    \item ``wide class of 2d objects ranging from \textbf{wires and ropes to hair}'' \cite{derouet2010stable}
    \item ``\textbf{ropes, chains, belts, cables, tendons, hair}, and other thin, curve-like physical objects'' \cite{sueda2011large}
    \item ``one-dimensional elastic \textbf{rods and hair}'' \cite{batty2012discrete}
    \item ``\textbf{chains} and chainmail, \textbf{threads}, knots, \textbf{hair, and necklaces}'' \cite{qu2021fast}
    \item ``beams, \textbf{cables, yarn, and hair}'' \cite{hafner2023design}
\end{itemize}
``Hair'' is imagined to be an entirely straight solid, such as ropes, chains, wires, belts, cables, and rods. The main model for hair simulation at both Walt Disney Animation \cite{thyng2017art} and Weta Digital \cite{lesser2022loki} is the \textit{Discrete Elastics Rods} \cite{Bergou:2008:DER} model, so this imaginary of ``straight as a rod'' carries over into films such as \textit{Moana} (2016).

One apparently missing paper is the curly hair system \cite{iben2013} Pixar developed for Princess Merida in the film \textit{Brave} (2012). The journal-length version of that paper was rejected by both \textit{SIGGRAPH} and \textit{ACM Transactions on Graphics}, and instead appeared in the secondary conference the \textit{Symposium on Computer Animation}. \cite{iben2013pc}

\paragraph{Hair Rendering}

We found 64 papers on rendering, 35 of which dealt directly with hair rendering algorithms. Of these, 21 papers (60\% of 35) showed exclusively straight and wavy hair, and 6 papers (17\% of 35) also included curly hair. There were 8 papers (23\% of 35) showing the full range of straight and wavy, curly, and textured hair, and all but one \cite{lombardi2019neural} were published after 2021. Within these rendering papers, we found that hair was initially abstracted as a translucent cylinder \cite{kajiya1989rendering}, and later refined to a cylinder covered in cuticles \cite{marschner2003light}. This conception of hair as a straight cylinder was reflected in 3 other papers on cloth rendering that directly compared hair to threads:
\begin{itemize}
    \item ``Like hair, textile fibers are long cylindrical structures made of dielectric material'' \cite{khungurn2015matching}
    \item ``unlike hair, threads have no consistent cuticle that displaces their specular reflection'' \cite{sadeghi2013practical}
    \item ``This is similar to previous work on hair rendering'' \cite{montazeri2020practical}
\end{itemize}

The special issue of \textit{TOG} on production renderers appeared again, this time including all 5 papers from the issue. The Arnold renderer \cite{georgiev2018arnold}, as well as Sony Imageworks' fork of Arnold \cite{kulla2018sony}, appeared in addition to the Pixar, Weta, and Disney renderers from \S\ref{sec:skin_optical}. All of the papers showed exclusively straight or wavy hair \cite{fascione2018manuka,christensen2018renderman,georgiev2018arnold,kulla2018sony,burley2018design}. A 2004 paper on the DreamWorks rendering system \cite{tabellion2004approximate} for \textit{Shrek 2} (2004) also appeared, and the only examples of hair were Prince Charming's blonde bob, and the fur on Puss In Boots.

\section{Discussion}
\label{sec:discussion}

\subsection{Binarization and Hierarchization}
\label{sec:binary}

\subsubsection{Binarization} Our results show binarization in the algorithmic conceptualization of skin. The physical substrate of skin is formulated as a high albedo, highly scattering material analogous to snow, milk, wax and marble: white skin. This conceptualization appears to gain momentum circa 2011 with the commercial success of \citet{jensen2001practical}. Prior to that paper, the imaginary surrounding skin rendering included darker (South Asian) \cite{marschner1999image}, and ``dark complexion'' skin \cite{hanrahan1993reflection}. After 2001, the possibility of humans with darker skin are acknowledged, but the algorithmic results put forward suggested that it was unnecessary to demonstrate that a technique can render equally convincing white and Black humans. The consistent, albeit unspoken, evidentiary benchmark is that results on white skin are acceptable synecdoche. Other shades of skin appear in large tables of disparate skin tones with no consistent standard for inclusion, aside from being darker than the white skin examples. We observe binarization into \textit{white} and \textit{non-white} skin.

We see similar binarization for hair. While for classification purposes we combined the categories for straight and wavy, the literature draws unambiguous analogies to ropes, wires, strands, and yarn: straight hair. Other hair types  appear as derivations from this canonical type, often as an inconsistently labeled table that conflates the categories of wavy and curly. The only unifying characteristic among the ``curly'' examples is that they are less straight than the straight hair examples. We observe binarization into \textit{straight} and \textit{non-straight} hair.

\subsubsection{Hierarchization} \citet{klein24data} observe that ``binaries are often hiding hierarchies.'' With skin, the non-white form is conceived to descend from the white form. With one exception \cite{williams1990performance}, every skin paper included an example of white skin. Many \textit{only} showed white skin, while claiming to apply generically to ``skin.''

The hierarchy is visible in Figure \ref{fig:layers}, where non-white skin receives a nod in the form of swatches and charts, but the final empirical validations only occur on white skin. These papers were accepted despite this gap in empiricism, suggesting that the hierarchy is a community-wide norm, not one researcher's idiosyncrasy. The same hierarchy implicitly appears in literature gaps. When high-albedo phenomena is shown to be less salient in high-melanin skin \cite{weyrich2006analysis}, alternative visual phenomena associated with dark skin (gloss) goes unexplored.

Similarly, almost every paper published at \textit{SIGGRAPH} includes straight hair. Non-straight hair is sufficiently subordinate that showing it is not necessary to claim that an algorithm applies to generic ``hair.'' The fungibility within the ``non-straight'' category is apparent in the images that get labeled as ``curly'' (Figure \ref{fig:curly_hair}). In at least two known instances \cite{iben2013,liftedCurls}, submitted papers that centered curly hair were rejected from \textit{SIGGRAPH}. In one instance, the failure to acknowledge the centrality of straight hair was a reviewer rationale for rejection \cite{kim2023sgp}.

\subsubsection{Racial Elasticity} \label{sec:elasticity} Racial categories are not fixed, but shrink and grow in order to preserve existing power structures \cite{cottom2019thick}. Privileging whiteness does not preclude representation of any other racial group \cite{bonilla2004bi,treitler2013ethnic}. Our survey showed that in the 1980s to 2000s, white-only facial capture papers were most common, while starting in 2008 and accelerating in the 2010s, white and East Asian-only papers became acceptable. In the 2020s, East Asian-only facial capture papers became consistently acceptable at \textit{SIGGRAPH}. Over the same time, no South Asian-only or Black-only papers appeared. This is consistent with scholarship on how the racial category "white" expands, and how its dominance can be buffered by other racial categories \cite{bonilla2004bi,hong2021minor}. The ongoing maintenance of a racial hierarchy privileging whiteness remains, to the exclusion of other (most notably Black or dark) racial groups. Under a theory of elasticity, ``white'' has expanded to include a light East Asian baseline,  following a period of economic growth in East Asia.  Pursuing a growing global market, multinational companies have circulated aspirational portrayals of Asian "lightness" \cite{glenn2008yearning, saraswati2020cosmopolitan, lee2019fashion}.

This inclusion has occurred within the existing assumptions of the computer graphics field, rather than shifting the status quo. This fits the pattern observed by \citet{lawrence2021} that these sorts of expansions are ``not being driven by inclusionary ideals, it is about accessing an emerging profit center.'' 
The profit incentives are also visible in the author affiliations of the \textit{Facial Capture} papers, where of the 26 with inclusive skin tones, 5 came from Disney Research, 4 from Meta, 3 from Shanghai Tech startup Deemos Technology, 2 from USC startup Pinsceen, and 1 paper each came from Tencent, Technicolor, Snap, Google, and Kuaishou Technology. However, the high albedo models for skin and straight rod assumptions for hair were not revisited. Darker skin and curlier hair were deemed worthy of biometric and economic \textit{capture}, but not of scientific \textit{study}.





\begin{figure}
\includegraphics[width=\textwidth]{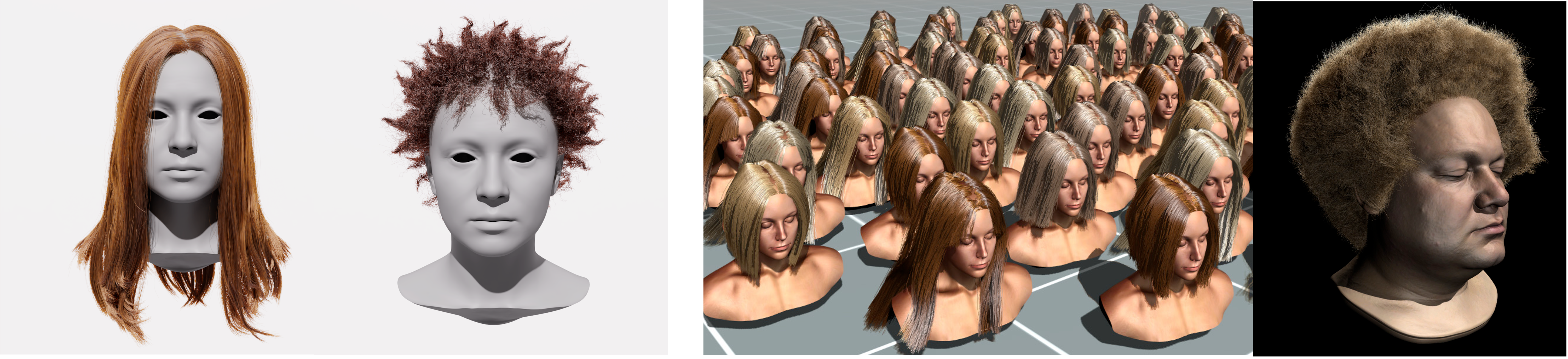}
\caption{Two tokenism examples. In the left pair \cite{daviet2023interactive}, a simulation technique is extensively validated on a straight hair example before presenting a perplexing noisy-haired example at the end. In the right pair \cite{bhokare2024real}, a rendering technique is shown scaling up to an army of straight-haired models, while a perplexing ``curly'' example appears at the end.}
\label{fig:mcdaniels}
\end{figure}

\subsubsection{Post-George Floyd Tokenism}
\label{sec:post_floyd}

Our review also found a surge in tokenism after the 2020 murder of George Floyd that coincided with the ensuing worldwide protests for racial justice. Many skin tone tables in facial capture papers finally expanded to include examples of Black skin the following year (18 of 25). The first instances of highly coiled hair (Walker Type 4 \cite{Walker1997}) appeared in 2023 \cite{hsu2023,zhou2023groomgen}, following significant internal \cite{kim2022countering,kim2022counteringBOF} and external \cite{kim2020racist} organizing that directly cited Floyd's murder. The only instance \cite{wu24curly} of a \textit{SIGGRAPH} paper treating highly coiled hair as the main research question came directly from this advocacy, i.e.~some of the organizers authored the paper (as well as this one).

Despite this apparent progress, in at least two post-2020 instances \cite{daviet2023interactive,bhokare2024real}, examples of highly coiled hair were shown that do not correspond to any form of human hair. Highly coiled hair is composed of tight helices, but in the given examples, they were modeled as undifferentiated noise layered atop straight hair. (Figure \ref{fig:mcdaniels}) These examples passed peer review as evidence that the techniques apply generically to ``hair,'' underscoring how ignorance of even the basic shape of highly coiled hair remains an entrenched community-wide norm.

\subsection{McDaniels and Durald Methods}

Binarization, hierarchization, and tokenism have antecedents in analog film and photography. \citet{dyer1997white} has described how ``[i]nnovation in the photographic media has generally taken ... the white face as the norm.'' This pattern repeated in \S\ref{sec:binary} with the standardization of high albedo materials as the optical substrate of skin. The influence of film and photography has been considerable in the history of \textit{SIGGRAPH} (Figure \ref{fig:industry}). Two of the most prolific labs from 1980-1990 were Lucasfilm and Pixar, with Disney becaming influential in the 2000s and 2010s. Microsoft's rise coincides with its release of the Xbox, and the largest current industrial player is Adobe, the company behind Photoshop.

\citet{dyer1997white} suggests that the underlying ideology of a technology is visible in its standard operating manuals. One prominent lighting manual that successfully spanned both the analog and digital eras is \textit{Painting With Light} by \citet{alton1949painting} in 1949. The phrase goes back further \cite{milner1930painting}, but Alton's work is the one referenced in the graphics literature \cite{pellacini2007lighting,lopez2010stylized,decoro2007stylized}. Alton displays the same hierarchization and tokenism in Chapter 5, ``The Hollywood Close-Up,'' when he exhaustively documents how to light a ``face,'' and every face is white. The only exception appears when Alton briefly notes that ``[t]here is a widespread belief that close-ups of colored people have to be overlit. Nothing can be farther from the truth. By lighting them normally, we get a bronze-like skin texture.'' The techniques used to light white actors can be transferred, without modification, to Black actors as well. Later scholarship has shown this to be false;  cinematographers experienced in lighting Black skin consistently note that larger, diffuse light sources are preferred in order to achieve better shaping on darker skin \cite{wiggum2021pixar,harding2017keeping}. The token example in \citet{alton1949painting} serves the rhetorical role of waving away concerns about the generalizability of a white skin technique.

The same flow appears in \textit{SIGGRAPH} papers when algorithms get exhaustively validated on white skin and straight hair, with a token example of dark skin or curly hair inserted at the end. In both cases (Figures \ref{fig:black_relighting} and \ref{fig:mcdaniels}), the results can be visually disappointing and missing key features. While \textit{an attempt was made} on a non-white example, whether it was successful is considered immaterial. While \textit{SIGGRAPH} papers are not directly imitating \citet{alton1949painting}, they are transferring widespread norms of whiteness, as documented in \citet{alton1949painting} and elsewhere, from the analog era into the digital.

As described in \S\ref{sec:intro}, we label techniques that deploy this rhetorical maneuver \textbf{McDaniels Methods}: \textit{A technique that is exhaustively validated on white skin and straight hair, and relegates all other types of skin and hair to a small set of flawed and underwhelming examples.} To reiterate, we chose this name because the actress Hattie McDaniel is the only Black subject in \citet{alton1949painting}, but in a reflection of the carelessness given to her depiction, her name is misspelled. Attention to Black representation has historically been an afterthought in cinematography; the evidence presented here suggests it has remained so in computer graphics research.

\newtext{We assign the label \textbf{Durald Methods} to the \textit{techniques that were co-created in close collaboration with the people being depicted.} We named this after Autumn Durald Arkapaw, the winner of the 2025 Academy Award for Best Cinematography for the film \textit{Sinners}. The lighting techniques for the film were designed in close collaboration with director Ryan Coogler to tell a story of racism and vampires in 1930s Mississippi \cite{Ryzik2026}}.

\subsection{Opportunities for Future Research}

We examined the 6864 papers from \textit{SIGGRAPH} and \textit{TOG} because they are the most influential conference and journal in computer graphics. Expanding our study would be valuable future work. We estimate from {\tt Crossref.org} that the journal corpus is $\sim$26602 articles across \textit{Computer Graphics Forum} (7104 articles), \textit{IEEE Transactions on Visualization and Computer Graphics} (7512), \textit{The Visual Computer} (5458), \textit{Computers \& Graphics} (5691) and \textit{Graphical Models} (837).

Stepping away from a norm of whiteness opens up a variety of graphics research directions. A relative absence of research into the complexities of melanin has already been observed in physics \cite{prescod2021disordered}, and while some graphics work also investigated its role in appearance, they are consistently McDaniels Methods \cite{aliaga2023hyper,chen2015hyperspectral}. There has been no \textit{SIGGRAPH} paper on the optical phenomena \textit{characteristic of high melanin skin}, such as the well-known ``blue undertones'' \cite{prescod2021disordered}. The topic is ripe for investigation. In addition, there has been no investigation into the appearance of gloss in high melanin skin. The representational conflict between a diffuse, ethereal ``glow'' and a hard, well-defined ``shine'' was examined by Dyer \cite{dyer1997white} in film and photography, where the ideal has always been ``white women ... bathed in and permeated by light ... in short, they glow.'' Tracing these visual norms back to painting, Thompson \cite{thompson2015shine} described how shine was used in ``the slave sublime,'' where ``slave traders actually greased the bodies of enslaved Africans, using sweet oil or greasy water `to make them shine,' as freedman Moses Roper described it ... to blind buyers, if you will, to the slave’s humanity.'' Glowing skin as the preferred mode of human depiction predates computer graphics, while investigating shine violates a hegemonic visual norm in European art. However, for scientists willing to step outside of this norm, \textit{developing a theory of high melanin shine} would advance the science of depiction.

Textured hair also offers multiple opportunities for investigation. The mechanics of \textit{why hair curls} is still poorly understood, as the hypothesis that it was due to the hair's elliptical cross-section was recently shown to be correlative, not causative \cite{wortmann2020hair}. A better understanding of where textured hair comes from would even have implications outside of computer graphics, in the same way that botanists have identified multiple mechanisms for tendril coiling in plants \cite{farhan2023artificial}. The spontaneous self-organization exhibited in textured hair also remains poorly understood. However, understanding \textit{how thin structures self-order} is of interest in soft-matter physics \cite{guerra2023self} and has applications such as DNA packing into cell nuclei \cite{kindt2001dna}. Detailed understanding of this phenomenon could potentially have wide impacts.

More broadly, the way forward is to reject the false promise of colorblind research and instead work to close heretofore neglected algorithmic gaps. \newtext{Doing so will require sustained engagement with practitioners skilled in the art of depicting Black skin and hair, and a commitment to Durald Methods over McDaniels Methods. We know of at least one such instance, involving authors from this paper, where graphics researchers entered into a years-long collaboration with an artist who specializes in the digital depiction of Black hair. A sustained conversation on the aesthetics of Black hair, its characteristic shapes and motions, and what features are considered culturally desirable, yielded the first paper on textured hair to ever appear at \textit{SIGGRAPH}. \cite{wu24curly} In a reflection of the public appetite for such Durald Methods, the work went on to garner widespread press coverage. \cite{ortega2024,gant2024,Rascoe2025,Bambury2025,Oladipo2025}
We hope that by identifying these positive and negative patterns, this current work can spur similar advances in the science of depiction.}
\section*{Generative AI Usage Statement} No generative AI was used in the preparation of this submission.

\section*{Positionality Statement}

Our research team comprises scholars from multiple disciplines, including computer science, information science, sociology, and American studies. Across these fields, our work is informed by critical science and technology studies, with particular attention to bias, race, and the social implications of computational systems. Our prior research has examined labor relations in AI development, racial boundary making in the U.S.~and Asia, and includes dozens of technical publications in computer graphics, including the \textit{SIGGRAPH} and \textit{Transactions on Graphics} venues examined here.

We conducted this study from our positions as researchers at Yale University, a well-resourced elite institution in the United States. The team has different backgrounds; one team member is an immigrant from Colombia working in the United States; another identifies as South Asian American, and the others as Korean American and White American. These institutional, disciplinary, and personal positions shaped how we approached the research questions, selected and interpreted methods, and understood the broader significance of the subject under study. We share these positions not to suggest that particular experiences are representative of the groups mentioned, but 1) to showcase that we do not all come from the position and 2) because our experiences in these positions contribute to our team’s collective racial commonsense understandings, which we implicitly drew upon when making skin and racial classifications in the study. 

There is no objective way of measuring difference, and we characterize our approach as subjective. Every ``objective'' classification system for skin is rooted in particular ideas about racial hierarchies, including in some cases eugenics and Nazism, as we note in \S\ref{sec:background} of the paper. Accordingly, as researchers, we eschewed these classification systems and instead applied blunt, high-level binary classifications: of skin as white/non-white and hair as straight/non-straight. Work in science and technology studies suggest that standard and classification efforts are never neutral, so we recognize that our classification efforts are necessarily subjective and not neutral either. We chose a classification strategy that allows us to open a broader conversation about race, while recognizing that the binary categories are too limited to do more than start that conversation. 

\section*{Acknowledgments}

This work was supported by NSF IIS-2132280, the Bungie Foundation, the Teng and Han Family Fund, and the Social Science Research Council’s Just Tech Program, with funds provided by the Ford Foundation, MacArthur Foundation, and Surdna Foundation. The views, findings, conclusions, or recommendations expressed in this paper are those of the authors and do not necessarily represent those of the International Development Research Centre, the Social Science Research Council, Ford Foundation, MacArthur Foundation, and the Surdna Foundation or their governing boards. The authors also would like to thank Lauren F.~Klein for early discussions and pointers to other field-level audits. 
\bibliographystyle{ACM-Reference-Format}
\bibliography{sample-base}

\appendix

\section{Methods}


\subsection{Data Processing}
\label{appen:data_process}

Through an agreement with our university library, we have been able to obtain the full text and metadata of every {\em SIGGRAPH}, \textit{SIGGRAPH Asia}, and \textit{ACM Transactions on Graphics} article from 1974 to 2024 from the ACM Digital Library. The data consisted of DTD BITS and JATS paper texts, programs, authors, affiliations and CCS classifications, among other data, for all papers in every issue of \textit{ACM Transactions on Graphics} and the \textit{SIGGRAPH} proceedings. Before analyzing the data, we performed three stages of manual data cleaning and labeling, which we refer to as \textit{Splitting}, \textit{Clustering}, and \textit{Classifying}.

\paragraph{Splitting}
The dataset schema allows for multiple listed affiliations per author, but we found that roughly one-fifth of affiliations were listed as a single entry despite referring to two or more institutions (e.g. \textit{``Company A and University B''}). These compound affiliation strings use no consistent delimiter and are often indistinguishable from non-compound affiliations without institution-specific knowledge, precluding automated splitting.

Instead, we performed splitting with a two-stage semi-automated approach. First, the 6,610 unique affiliation strings from the papers dataset were categorized as either compound or non-compound. Each of the compound affiliation strings was then split by the human reviewer into its component parts, using an interface that highlighted likely conjunctions but that allowed the human labeler to split the string into arbitrary elements. All affiliation strings were split by the same human reviewer.

\paragraph{Clustering}
\label{sec:clustering}
Splitting reduced the total number of unique affiliation strings from 6,610 to 5,913.
As many of these referred to the same institution under different names (e.g., ``\textit{EPFL}" and ``\textit{EPF Lausanne}"), we sought to cluster affiliation strings that refer to individual institutions. As an exhaustive performance of this task would require $N^2>3\mathrm{e}7$ human string comparisons, several heuristics were used to produce approximate clusterings.

For each unique affiliation string, a human labeler was shown a list of key collisions using three different text collision-finding algorithms from the OpenRefine project (\textsc{Fingerprint}, \textsc{NgramFingerprint}, and \textsc{Metaphone3} keyers). The human labeler also performed text searches of all affiliation records to select other results for clustering. After this process was completed, any overlapping clusters were manually reviewed and either combined or disentangled to produce a set of 1,653 disjoint clusters of affiliations.


\paragraph{Classifying Country Codes}
\label{sec:geotag}
For each individual affiliation string within the clusters, we then determine the approximate geographic location. Each post-split affiliation string was first geocoded using the Google Maps Geocoding API to assign an ISO 3166-1 country code. Each affiliation/location tag was then examined by a human reviewer, who verified the geocoding and made corrections where necessary. The Geocoding API returned a result for roughly $90\%$ of the affiliations, and our human review pass found an approximately $2\%$ error rate among the returned country codes. All reviewers were graphics researchers who had familiarity with the field and the publishing institutions. The results of this data analysis was used to generate Figure \ref{fig:industry}.

\section{Results}
\label{appen:results}

\subsection{Skin Papers}
\label{append:skin}

\subsubsection{Skin as an optical medium}
\label{append:skin_optical}

The \textit{Manuka} \cite{fascione2018manuka} paper describes the technical choices needed to ``render high-quality skin by path-traced subsurface scattering.'' The paper tags two images as ``skin'' renderings. One is of \textit{Digital Emily} \cite{alexander2009digital}, a photorealistic rendering of a white actress, and one of pale, iridescent aliens from the film \textit{Valerian and the City of Thousand Planets} (2017).

The \textit{RenderMan} \cite{christensen2018renderman} paper describes its strategy for rendering ``skin and other translucent materials'' using subsurface scattering and shows images of Arnold Schwarzenegger from \textit{Terminator Genisys} (2015), Peter Cushing from \textit{Rogue One: A Star Wars Story} (2016) and Sean Young from \textit{Blade Runner: 2049} (2017). 
    
The \textit{Hyperion} paper \cite{burley2018design} draws an explicit equivalence between skin and snow in the section \textit{Unifying Subsurface Scattering, from Snow to Skin}. The skin algorithm started as experiments for the snow monster Marshmallow in the short \textit{Frozen Fever} (2015) and the snowscapes in \textit{Olaf's Frozen Adventure} (2017). With the success of these experiments, ``path-traced subsurface scattering is now being used in all current productions on all materials from snow to skin.'' The one example in the paper labeled as ``skin'' shows Ralph and Vanellope from \textit{Ralph Breaks the Internet: Wreck-It Ralph 2} (2018). Both characters are white. Darker-skinned characters from \textit{Moana} (2016) also appear in the paper, but as showcases for the engine's denoising and ocean rendering capabilities. 

\subsubsection{Skin as a layered material}

\begin{itemize}
    \item A 2003 paper \cite{tsumura2003image} on hemoglobin and melanin only demonstrates its algorithm on a pale-skinned East Asian faces. The authors are all based in Japan.
    \item A 2005 paper \cite{donner2005light} on multi-layered translucent materials models skin using 6 different layers, and ends with a single rendering of a white face.
    \item A 2008 paper \cite{donner2008layered} proposes a ``layered, heterogenous''  model for skin, and renders test swatches of skin that include one instance of ``African, skin type V'', in reference to the Fitzpatrick skin typing system \cite{fitzpatrick1975soleil} from \S\ref{sec:skin_methods}. The final showcase renders of human hands and an ear are all of white skin.
    \item A 2010 paper \cite{jimenez2010practical} on dynamic facial color only demonstrates results on a white model. The data was collected from ``one Caucasian female, 33 years old; three Caucasian males, 26, 33, and 35 years old''. A ``skin color lookup texture'' is proposed to alter the model's melanin, but is never applied on a facial render.
    \item A 2015 paper \cite{chen2015hyperspectral} only showcases results on white models. The possibility of darker skin due to different melanin levels is mentioned but not demonstrated.
    \item A 2015 paper \cite{nagano2015skin} on capturing the microstructure of skin, e.g.~the forehead wrinkles that form under frowning, only shows results on white skin.
\end{itemize}

\subsubsection{Skin as a Human Face}
\label{sec:skin_face}

\paragraph{Facial Capture} Of the 67 papers we examined:
\begin{itemize}
    \item 18 contained only whites. \cite{waters1987muscle,guenter1998making,pighin1998synthesizing,bradley2010high,beeler2011high,beeler2012coupled,valgaerts2012lightweight,bickel2012physical,bermano2013augmenting,garrido2013reconstructing,bermano2014facial,beeler2014rigid,ichim2015dynamic,thies2015real,wu2016anatomically,zoss2018empirical,thies2018facevr,aneja2023clipface} The papers spanned the 1980s \cite{waters1987muscle}, 1990s \cite{guenter1998making}, and 2010s \cite{thies2015real}. The most recent appeared in 2023. \cite{aneja2023clipface}
    \item 10 contained only whites and East Asians. \cite{ma2008facial,ghosh2011multiview,shi2014automatic,fyffe2014driving,cao2015real,olszewski2016high,cao2016real,zoss2019accurate,zhu2024fabrig,chen2024monogaussianavatar} Of these, 7 were from the 2010s, and the most recent from 2024. \cite{chen2024monogaussianavatar}.
    \item 3 contained only whites and South Asians, and appeared in 2018, 2022 and 2023. \cite{wang2023neural,gotardo2018practical,winberg2022facial} One 2007 paper \cite{bickel2007multi} contained an example where our labellers could not agree on a South Asian or Black label. One conflating factor is that the paper was written by European authors, implying that the subject was also European. Thus, the North American labellers had difficulty parsing the European grooming aesthetics.
    \item 3 contained only whites and East Asians, plus Barack Obama. \cite{garrido2016reconstruction,egger20203d,gao2022reconstructing}
    \item 3 contained only East Asians \cite{bao2021high, ma2022neural,zhang2022video}, and all where published in the 2020s.
    \item 2 contained whites, East Asians, and South Asians \cite{moser2021semi,yang2024learning}, and both were published in the 2020s.
    \item 1 contained solely Black examples \cite{williams1990performance} and was published in 1990 by a white researcher.
    \item 26 contained skin tones ranging white to Black. All appeared after 2010 and 85\% after 2020. \cite{beeler2010high,hu2017avatar,nagano2018pagan,cao2018stabilized,li2020dynamic,riviere2020single,schwartz2020eyes,feng2021learning,cao2021real,chandran2021rendering,cao2022authentic,wang2022morf,zoss2022production,liu2022rapid,qiu2022sculptor,mendiratta2023avatarstudio,duan2023bakedavatar,zhang2023dreamface,zhang2023hack,kirschstein2023nersemble,yang2023towards,buehler2024cafca,kirschstein2024gghead,teotia2024hq3davatar,baert2024spark,bai2024universal}
\end{itemize}

\paragraph{Facial Reflectance and Relighting}
The initial 2000 work of \citet{debevec2000acquiring} on the topic prominently features a dark skinned model. Its proposal to use a metallic model \cite{torrance1967theory} to model skin glossiness was later adopted by \citet{weyrich2006analysis}. Immediately following the publication of the influential 2001 subsurface scattering paper \cite{jensen2001practical}, reflectance field papers from Debevec in 2002 and 2005 only featured white skinned final renders \cite{debevec2002lighting,wenger2005performance} (though \cite{debevec2002lighting} shows Black skin in a didactic figure) before later including East Asian \cite{peers2007post} and then South Asian and Black \cite{ghosh2008practical} renders in 2007 and 2008. These coincided with the addition of East and South Asians on the author lists.
Overall, we found:
\begin{itemize}
    \item 3 papers feature white-only final renders, the most recent from 2024. \cite{meka2019deep,li2024uravatar,wenger2005performance} 1 additional paper includes a Latina woman \cite{guo2019relightables}, though the highly diffuse render suggests white aesthetics.
    \item 2 papers show whites and East Asians \cite{peers2007post,yoshihiro2018relighting}, and while one of these \cite{yoshihiro2018relighting} shows a Black example in their training set, there are no Black examples in their final evaluation. 1 additional paper \cite{nishino2004eyes} shows East and South Asians, and 1 additional paper features white, East Asians, and South Asians. \cite{bharadwaj2023flare}
    \item 12 papers show final renders across a range of skin tones. One is from 2000 \cite{debevec2000acquiring}, two are from the 2010s \cite{legendre2016practical,yamaguchi2018high}, and the rest (75\%) are from the 2020s \cite{meka2020deep,sun2020light,bi2021deep,yeh2022learning,sarkar2023litnerf,he2024diffrelight,rao2024lite2relight,yang2024vrmm,zhang2021neural}. 
\end{itemize}







\subsection{Hair Results}
\label{append:hair}

\paragraph{Hair Geometry}

Within the category of curly hair, we observed subcategories of \textit{European locks} \cite{Darke24}, which are straight at the root but curly near the tip, and \textit{pageant hair} \cite{knighter2018}, which is straight hair with a high-radius, low-frequency twist artificially inserted near the tip. Pageant hair style is politically charged in the United States, where it is also called \textit{Big Washington Hair} \cite{kahn2025} and \textit{Republican Hair} \cite{defalco2025}. We found the following across 61 hair geometry papers:
\begin{itemize}
    \item 17 papers present results exclusively on straight hair. Of these, one reports failing on ``frizzy'' hair \cite{lawrence2021project}, and two showed textured hair in the training set but not the final results \cite{kirschstein2023nersemble,buehler2024cafca}. 
    \item 2 papers present results on straight hair, plus Barack Obama. \cite{duan2023bakedavatar,xu2023avatarmav}
    \item 10 papers present results on straight and wavy hair, with 2 from 2024. \cite{paris2004capture,paris2008hair,xu2014dynamic,chai2016auto,zhang17data,liang18video,bharadwaj2023flare,gao2023sketchfacenerf,kirschstein2024gghead,li2024uravatar}
    \item 2 papers present results on straight and wavy hair \cite{echevarria2014capturing,hu2014robust}, as well as European locks.
    \item 5 papers \cite{hu2015single,saito20183d,chai2012single,kuang22deep,Zheng24towards} present results on straight and wavy hair, plus pageant hair. Of these, 3 papers were from China-based researchers \cite{chai2012single,kuang22deep,Zheng24towards}, and 2 from the U.S. \cite{hu2015single,saito20183d}

    \item 2 papers deal with braids \cite{hu2014capturing,xiao2021sketchhairsalon}, which exclusively means straight (predominantly blonde) hair braided into ponytails or pigtails. Textured hair styles such as box braids, cornrows, or locs did not appear.
    \item 19 papers show examples of straight, wavy, and curly hair, but the definition of ``curly'' is \textit{highly} inconsistent, and often corresponded to lightly wavy hair. \cite{kim02interactive,wei2005modeling,wang2009example,yuksel09hair,herrera2012lighting,luo2013structure,chai2015high,zhang18model,Shen23ct,hsu24real,Reshetov24modeling,lombardi2021mixture,cao2022authentic,sun2022ide,bai2024universal,giebenhain2024npga,teotia2024hq3davatar,yang2024vrmm,zhao2024invertavatar}
    \item 2 papers show textured hair as failure cases, one involving box braids  \cite{saito20183d} and the other ``frizzy'' hair \cite{chai2015high}.
    \item 2 papers contain textured hair, one from 2023 \cite{zhou2023groomgen} and the other from 2024. \cite{wu24curly}
\end{itemize}


\paragraph{Hair Simulation}

We found the following across 35 hair simulation papers:
\begin{itemize}
    \item 15 papers exclusively showed straight hair. They variously dealt with hair, thin solids, and multiphysics simulations (interaction of cloth, rods, solid volumes, and fluids). \cite{mcadams2009,anjyo1992,bergou2010discrete,bertails2011nonsmooth,daviet2011hybrid,derouet2013inverse,ma2013dynamic,jiang2017anisotropic,michels2017stiffly,fei2019multi,chen2020moving,fei2021revisiting,li2021codimensional,lesser2022loki,wang2023fast}
    \item 4 papers showed exclusively straight and wavy hair. \cite{chai2014reduced,kaufman14,fei2017multi,hsu2022general}
    \item 2 papers showed exclusively straight and curly hair. \cite{Selle08,daviet2020simple}
    \item 2 paper showed straight, wavy and curly hair. \cite{superhelices,daviet2023interactive}
    \item 1 paper showed exclusively European locks. \cite{casati2013super}
    \item 2 papers from 2023 and 2024 \cite{hsu2023,crespel2024} showed straight, wavy, curly, and textured hair.
\end{itemize}

\paragraph{Hair Rendering}
We found the following hair categories across 64 hair rendering papers:
\begin{itemize}
    \item 11 papers \cite{kajiya1989rendering,marschner2003light,moon2006simulating,ragan2007lightspeed,jakob2009capturing,zhou2009renderants,sadeghi2010artist,barringer2012high,heitz2015sggx,sarkar2023litnerf} showing exclusively straight hair. One of these papers \cite{heitz2015sggx} shows a ``hairball'', (see below), but it is a ball of straight hair, not the noisy hair ball that other algorithms are tested on. Another \cite{jakob2009capturing} labels an example as ``wavy'', but labelers agreed that under the current scheme, it should be classified as ``straight''. The example showed a half-period oscillation.
    \item 6 papers \cite{zinke2008dual,o2010optical,ren2010interactive,xu2011interactive,khungurn2017azimuthal,xia2023practical} on straight and wavy hair, and 4 \cite{wenger2005performance,xu2019adversarial,zheng2021ensemble,back2022self} on wavy hair only
    \item 2 papers \cite{moon2008efficient,xia2020wave} on straight and curly hair, and 1 paper \cite{locovic2000deep} on curly hair only
    \item 3 papers \cite{zinke2009practical,wang2020single,mullia2024rna} on straight, wavy, and curly hair. One paper \cite{wang2020single} had an instance (Figure 10, bottom row) of potentially textured hair, but the hair was deliberately covered with a wrapping to exclude it from evaluation. Only small portions peeked out of the covering that might qualify as textured.
    \item 8 papers \cite{lombardi2019neural,bi2021deep,pandey2021total,balbao2022biologically,yeh2022learning,wang2023adaptive,bhokare2024real,he2024diffrelight} that show straight, wavy, curly, and textured hair. All but one \cite{lombardi2019neural} were published after 2021.
\end{itemize}
There were 9 papers \cite{perlin1989hypertexture,banks1994illumination,fleischer1995cellular,matusik2002image,cook2007stochastic,velazquez2009automatic,hou2010micropolygon,moon2010cache,gupta2023mcnerf} showing renderings of ``fur'', referring either to very short hairs, or hairs on an animal. An additional 5 papers \cite{goldman1997fake,yan2015physically,yan2017bssrdf,yan2017efficient,zhu2022practical} dealt specifically with the problem of ``fur'' rendering, whose appearance is considered distinct from human hair.  Another 5 papers \cite{havsan2006direct,lefohn2007resolution,didyk2010apparent,barringer2013a4,carra2019scenegit} showed a standard ``hairball'' example, which appears to be straight hair with noise applied.

\end{document}